\documentclass[preprint,12pt,numbers,sort,compress]{elsarticle}

\usepackage{amssymb}
\usepackage{amsmath}
\usepackage{lineno}
\journal{Environmental Pollution}

\begin{document}

\begin{frontmatter}



\title{Terahertz prototype for air pollutants detection}


\author[1]{Candida Moffa \corref{cor1}}
\author[1,2]{Alessandro Curcio}
\author[1]{Camilla Merola}
\author[1]{Vittoria Maria Orsini}
\author[1,2]{Daniele Francescone}
\author[1]{Fernando Jr. Piamonte Magboo}
\author[1]{Marco Magi}
\author[1]{Massimiliano Coppola}
\author[1,2]{Lucia Giuliano}
\author[1,2]{Mauro Migliorati}
\author[1]{Giuseppe Zollo}
\author[3]{Massimo Reverberi}
\author[1]{Leonardo Mattiello}
\author[1,2]{Massimo Petrarca \corref{cor1}}

\cortext[cor1]{Corresponding authors: candida.moffa@uniroma1.it; massimo.petrarca@uniroma1.it}

\affiliation[1]{organization={Department of Basic and Applied Sciences for Engineering (SBAI), Sapienza, University of Rome},%Department and Organization
            addressline={Via Antonio Scarpa, 16}, 
            city={Rome},
            postcode={00161}, 
            country={Italy}}
\affiliation[2]{organization={Roma1-INFN},%Department and Organization
            addressline={Piazzale Aldo Moro, 2}, 
            city={Rome},
            postcode={00185}, 
            country={Italy}}
\affiliation[3]{organization={Department of Environmental Biology (DBA), Sapienza, University of Rome},%Department and Organization
            addressline={Piazzale Aldo Moro, 2}, 
            city={Rome},
            postcode={00185}, 
            country={Italy}}

%% Abstract
\begin{abstract}
In this work, we propose a prototype set-up exploiting terahertz time-domain spectroscopy (THz-TDS) to investigate gaseous compounds. The system is portable and allows to perform remote measurements. We used the prototype to characterise for the first time in literature over a broad THz range, pure dichloromethane and chloroform, two pollutants known as very short-lived substances (VSLS) that strongly contribute to ozone depletion. The THz range allows selectively detecting their absorption lines related to the rotational molecular motion for which we also present the theoretical confirmation.
Then, we investigate the optical response of a multi-component mixture achieved with the two aforementioned chlorine-based compounds mixed with two widely distributed volatile pollutants (acetone and methanol). For these first measurements, we developed the set-up specifically for \textit{laboratory condition} in which the substances are directly injected into the gas-cell circuit. Finally, we modified the prototype to ensure that the ambient atmosphere is drawn directly into the gas cell via a long pipe and a suction system opportunely developed. 
The analysis of the mixtures in both \textit{laboratory} and \textit{in-field conditions} demonstrates that the prototype together with the approach employed in this work can simultaneously identify and quantify single components in the atmosphere.
The results obtained open up new possibilities for the development and applications of an efficient portable THz-based sensor for the remote detection of multi-component environmental contaminants.
\end{abstract}

\begin{graphicalabstract}
\includegraphics[width=\linewidth]{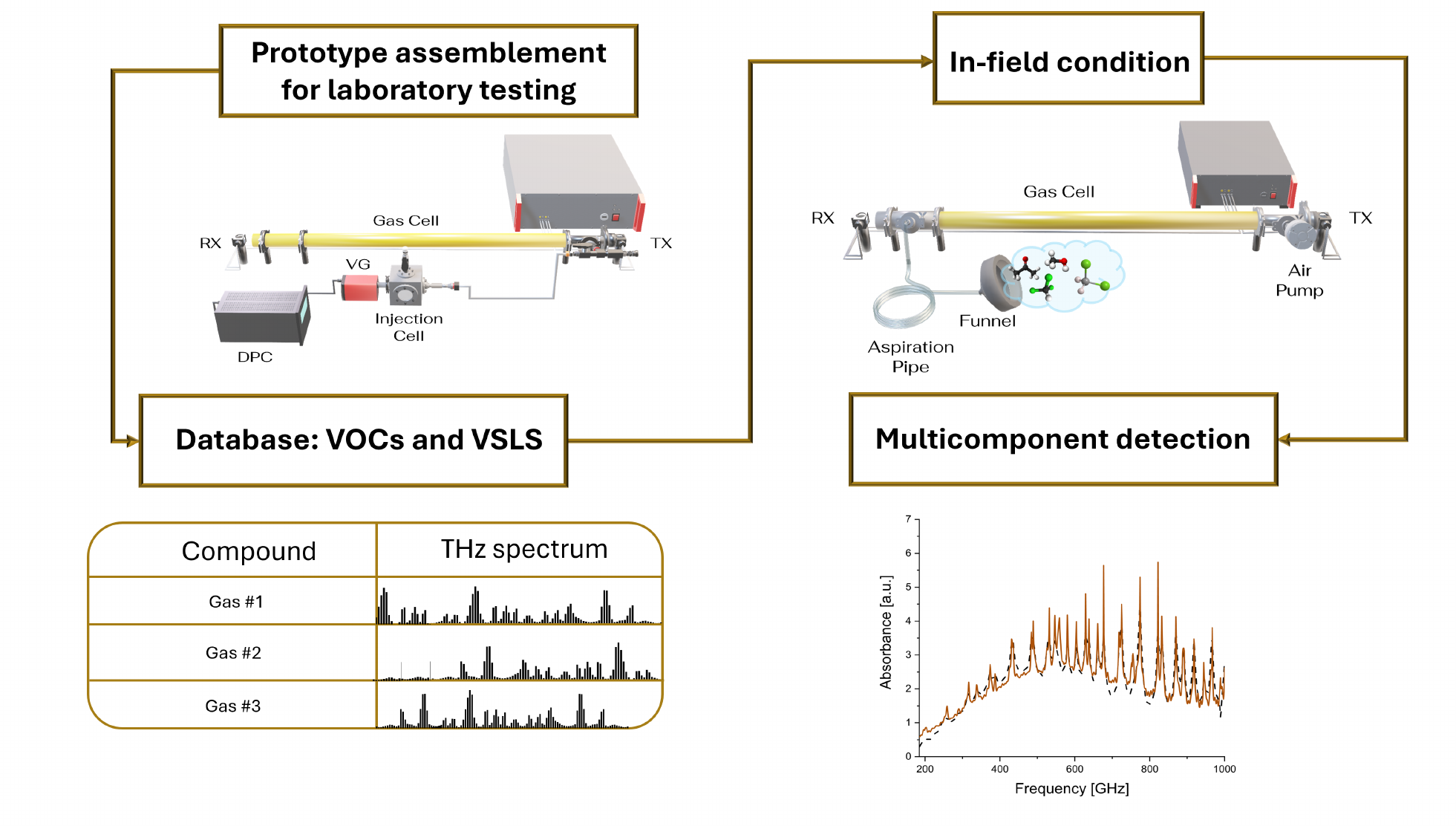}
\end{graphicalabstract}

\begin{highlights}

\item This study presents the first characterization of pure dichloromethane and pure chloroform in the terahertz (THz) spectral range up to 4.5 THz, offering a novel spectral fingerprint for these Very Short-Lived Substances (VSLS) which are significant contributors to ozone depletion.

\item A portable THz spectroscopy prototype was developed, capable of remote and in-situ detection of multiple air pollutants, including VOCs and VSLS. This system enhances real-time environmental monitoring by enabling measurements in both controlled laboratory and field-like atmospheric conditions.

\item The integration of a multi-absorber spectral fitting model with the THz system allows for accurate simultaneous identification and quantification of individual components within complex gas mixtures.

\item The prototype was successfully adapted for in-field measurements by integrating a custom-engineered suction system and a five-meter intake line, enabling direct sampling of ambient air from a controlled environment. Spectral data closely match the multiple absorbers model validating the system’s capability for robust, simultaneous identification and quantification of individual atmospheric constituents, underscoring its potential for advanced environmental monitoring applications.

\end{highlights}

%% Keywords

\end{frontmatter}



\section{Introduction}
The increasing presence of pollutants stemming from anthropogenic activities made their detection a major topic in the last decades.
Their identification holds significance for the safeguarding of ecological integrity since they can pose substantial threats to ecosystems, air and water quality, biodiversity, and human health.
Employing advanced monitoring techniques and analytical methods can help identify specific compounds, track their source pathways, and make efforts towards the implementation of effective mitigation strategies and the evaluation of the impact of human activities. 
For this reason, new efforts have been made to develop and optimise remotely controlled systems which can selectively identify these substances such as Volatile Organic Compounds (VOCs) \cite{Sekhar2014, DArco2022, Foronda2023, Shooshtari2023, Hajivand2024}, Green House Gases (GHGs) \cite{Abdelraheem2023, Rahman2023, DArco2022} and very short‐lived substances (VSLS) \cite{Wu2023, Sekhar2014}.

Specifically, VSLS are defined as trace gases characterised by brief atmospheric lifetimes ($<$ 6 months) that are of interest as environmental pollutants since simulations suggest a significant contribution to the ozone layer depletion in the lower stratosphere \cite{Hossaini2017, Hossaini2019, Laube2008, Fuhlbrugge2016, Ziska2017, Jia2019, Maas2020, Tegtmeier2020}. 
Among the halogen-containing substances, chlorine sources represent one of the major contributors to this phenomenon. Thus, their monitoring in the stratospheric layer has become of critical importance over the last two decades \cite{Hossaini2019}. 
The emission of chlorinated VSLS including dichloromethane ($CH_{2}Cl_{2}$, DCM) \cite{Dekant2021, Rice2021, Butz2006} and chloroform ($CHCl_3$) \cite{Sekar2021} is originated predominantly from anthropogenic activities \cite{Butz2006} although natural sources also contribute \cite{Butz2006} and they represent some of the most frequently encountered species \cite{Deng2020, Mcculloch2003}.

DCM is a high-production volume chemical \cite{Dekant2021} produced by natural 
processes in seawater and soils, and by biomass burning \cite{Dekant2021}. 
It is used as a solvent in synthesis, extraction, and purification purposes (i.e., essential oils, food processing, pharmaceuticals) and in the manufacture of plastics, as blowing agents in polyurethane foams, paint strippers, and special adhesives \cite{Shestakova2013, Dekant2021, Schlosser2015}.
A specific use which increased in recent years is in the manufacture of hydrofluorocarbons (i.e., difluoromethane) used as replacements for chlorofluorocarbons (CFCs) and hydrochlorofluorocarbons (HCFCs) \cite{Hossaini2017}.
Around 80-90\% of the produced DCM is released in the atmosphere \cite{Shestakova2013, Deng2020}.
Despite its insufficient reactiveness in the atmosphere to be classified as a “significant photochemically active volatile organic compound”  \cite{Dekant2021}, recent observations show a rapid increase in the tropospheric abundance of DCM \cite{Hossaini2017}.
Therefore, its contribution to the formation of tropospheric ozone and photochemical smog, makes its control a major environmental concern \cite{Deng2020}.
High indoor concentrations of DCM have been reported also in occupational settings \cite{Schlosser2015} and it poses a threat to human health since it is categorized as a category B2 probable human carcinogen \cite{Shestakova2013, Dekant2021}.

Chloroform ($CHCl_3$) has a half-life in the atmosphere in the range between 55 and 620 days and represents a threat to the ozone layer \cite{Sekar2021}.
Its production is also associated with natural sources \cite{Sekar2021, Mcculloch2003}, however in the last two decades increasing concentrations in the atmosphere have been connected to global emissions from anthropogenic activities \cite{Trudinger2004, Worton2006, Sekar2021}.
In particular, $CHCl_3$ is emitted from the natural and induced anaerobic fermentation in biogas generation, it is involved in the manufacturing of dyes, pesticides, insecticides and fumigants.
It is also widely used in pharmaceutical, fertilizer, polymers, paper and solvent manufacturing industries \cite{Sekar2021, Mcculloch2003}.
Moreover, it is one of the most encountered disinfection by-products in water treatments \cite{Sekar2021, Mcculloch2003}
Although it is not a strong mutagen, it is classified as a group B2 compound (a probable human carcinogen) \cite{Sekar2021, Mcculloch2003}.

In this context of gasses sensing, terahertz-based (THz) techniques showcased an increasing potential for their detection capabilities \cite{DArco2022, Tekawade2018}.
In fact, despite the several scientific and industrial domains in which THz spectroscopies have found applications (e.g. pharmaceutical, medical and biophysics studies through the non-invasive examination of molecules and the characterization of biological tissues \cite{Zhan2023, Chernomyrdin2023, Yang2023, Gezimati2023, Nourinovin2023, Wu2023};
for detecting concealed objects and harmful substances or drugs \cite{Ikari2023, Kumar2023, Singh2024, Witko2012}, to examine Cultural Heritage materials \cite{Kleist2019, Kleist2019(2), Moffa2024, Fukunaga2023, Fukunaga2024, Moffa2024(2)} and to monitor food quality and composition \cite{Dinovitser2017, Chen2022, Rawson2022}),
the versatility of terahertz spectroscopy underscores its significance as a powerful analytical tool with far-reaching implications also for detecting of environmental pollutants.
Indeed, THz radiation provides a selective identification of gas molecules since this range of frequencies results topical for detecting absorption lines of rotational molecular motion \cite{Lin2008, Vaks2020, DArco2022, Rice2021}. 
Furthermore, the THz range encompasses the low-frequency vibrational transitions of numerous molecules and thanks to the related low photon energy it cannot induce molecular ionisation nor combustion of flammable materials thus resulting in non-hazardous radiation to biological subjects \cite{Naftaly2015}.
Investigating gas molecules with THz radiation can also offer an increased selectivity compared to infrared transitions \cite{Rice2020} and, in addition, THz radiation is less affected by scattering since THz wavelengths are larger than the typical dimensions of airborne particulates (0.01-100 $\mu$m) \cite{Bigourd2006, Rice2021}.
However, despite the high potential of THz for gas-sensing, databases are still lacking the fingerprints of many substances of interest as VLSL in this spectral region; moreover, practical set-up for \textit{in-field} operation is still not well established or reported at these frequencies.

For the aforementioned reasons, in this work, we present a compact portable optimized set-up based on a terahertz time-domain spectroscopic system (THz-TDS) coupled to a gas cell for the investigation of gaseous compounds at a remote distance. To the best of the author’s knowledge, the proposed set-up combined with dedicated data analysis for the identification and quantification of gas substances represents a singular device for applications in real-case scenarios. 
We firstly used the proposed set-up to increase our THz dataset by studying the optical response of pure chloroform and pure dichloromethane that, to the best of the authors' knowledge, represent the first THz characterisation of the two VLSL halogenated compounds in the THz spectral region over a large bandwidth up to 4 THz. The characterization of these two compounds is presented alongside the relative theoretical explanation. 

Then, we study the possibility of exploiting the system to analyse and selectively monitor multi-component gaseous target \cite{Bigourd2006, Li2020, DArco2022} showing that the proposed scheme coupled with an analysis methodology based on the multiple absorbers approach for non-interacting compounds allows identifying and quantifying the single components in the gaseous mixture simultaneously. 
The first tests are performed in laboratory conditions, in which we directly inject the substances into the gas cell circuit where the mixture interact with the THz beam. The analyses of the THz spectrum by fitting it with the multiple absorbers approach gives a good matching between the experimental data and the fit.
Then, we moved on to the \textit{in-field configuration} by implementing the set-up with a pumping system for the aspiration into the gas cell of the compounds dispersed in the atmosphere at a distance of 5 meters from the gas cell.
Also in this case, the proposed set-up and methodology allow retrieving and quantifying the substances diluted in the atmosphere demonstrating its potential for in-situ applications.
%The obtained results evidenced that the laboratory system can be exploited to retrieve the presence of all the compounds.
%opening the possibility of developing a more compact and portable set-up for on-field measurements.

The presented results show a substantial leap in environmental monitoring and industrial applications by THz radiation, offering a reliable solution for remote compound detection. The ability to perform precise measurements in diverse locations without the constraints of a traditional laboratory set-up not only enhances efficiency but also broadens the scope of potential applications of THz spectroscopic systems.

\section{Results}

\subsection{Characterization of pure Dichloromethane and pure Chloroform}
\begin{figure}[htbp!]
    \centering
    \includegraphics[width=\linewidth]{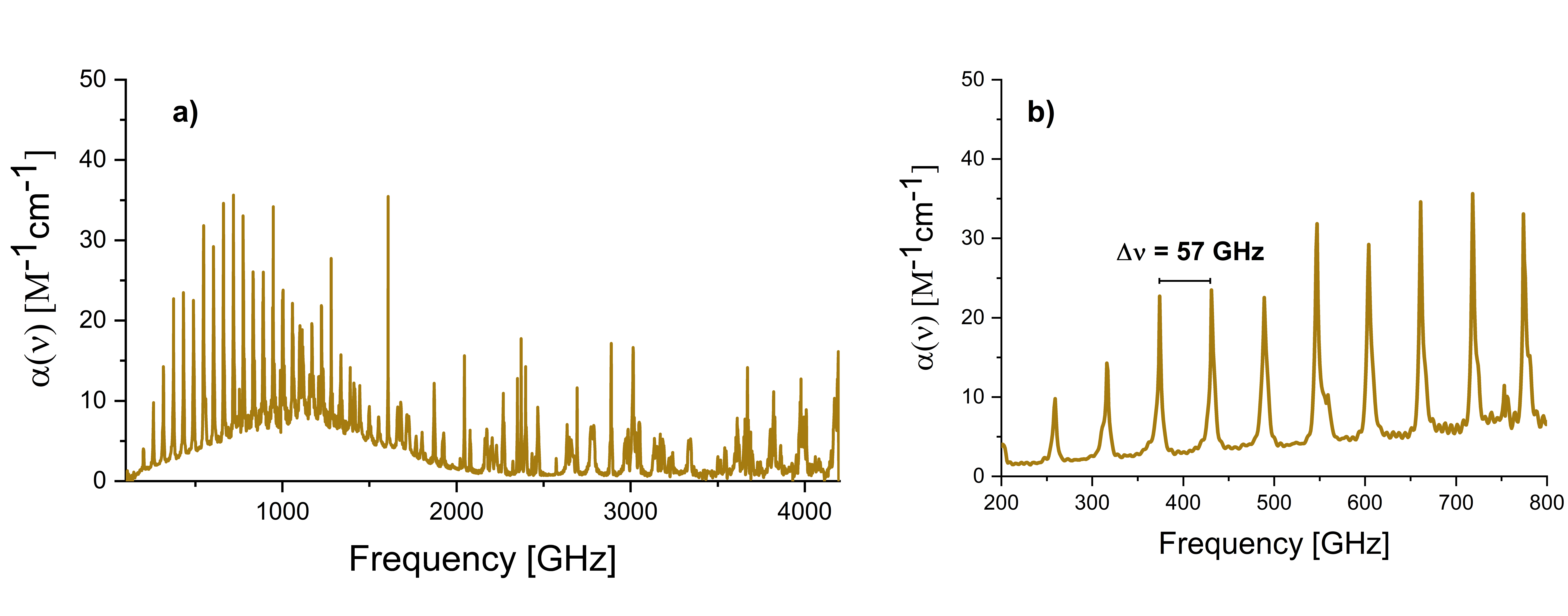}
    \caption{a) THz-TDS absorption spectrum of DCM (asymmetric molecule near the prolate limit with Ray’s parameter $\kappa$=-0.98) in the range 0.1-4.2 THz where rotational transitions of the Q-branch are spaced every $\Delta \nu$=57 GHz; b) THz-TDS absorption spectrum of DCM in the range 0.2-0.8 THz where the rotational transitions frequencies are reported.}
    \label{DCM_TDS}
\end{figure}
\textbf{Dichloromethane} (DCM) presents three isotopomers, the most occurring one (with a natural abundance of 56\%) is an asymmetric top molecule with rotational constants A=32.002 GHz, B=3.320 GHz, C=3.065 GHz \cite{Rice2021, Tullini1989}. 

Its Ray's asymmetry parameter, $\kappa$=-0.98, provides information on its geometry that can be considered near to the prolate limit (since for a prolate molecule $\kappa$=-1) \cite{Cooke2013}.
The permitted changes in the rotational quantum number $J$ are $\Delta J=-1;0;+1$.
At low frequencies DCM spectrum (Figure \ref{DCM_TDS}) shows a frequency spacing that approaches 2(A-B) \cite{Brown1969} which for DCM corresponds to rotational transitions spaced every 57.464 GHz which represents Q-branch transitions since $\Delta J=0$ (where $J$ is the total rotational angular momentum quantum number) \cite{Cooke2013}. This spacing decreases with increasing frequencies due to centrifugal distortion effects.
The experimental data obtained with terahertz spectroscopy highlight that the most evident absorption features are the perpendicular bands \cite{Brown1969}, in which the electric moment oscillates perpendicular to the principal axis.
The strong absorption features cover the entire range up to 2.3 THz and are associated with molecular rotation for $\Delta$J=0 from $J'$=1 $J''$=0 up to $J'$=84 and $J''$=83 where $K$ assumes values between $-J$ to $+J$ (where $K$ is the quantum number that represents the portion of the total angular momentum that lies along the symmetry axis, thus is always equal to or less than $J$). 
%(Fig. \ref{DCM_time}).
%\begin{figure}[htbp]
 %   \centering 
  %  \includegraphics[width=\linewidth]{DCM_time.jpg}
   % \caption{DCM's pulses sequences in time-domain in a 200 ps range.}
    %\label{DCM_time}
%\end{figure}
%The temporal waveform exhibits a series of periodically located THz echoes also known as Free Induction Decay (FID) signals consisting of a signal sequence separated by a temporal delay multiples of the inverse of the rotational line spacing. In agreement with this behaviour, the FID signals occur in this case approximately every 17 ps, corresponding to a quantum rotational revival period of $T_{rev}=[2(A-B)]^{-1}$ \cite{Lu2016}. The FID echoes decay because of relaxation, interference, and propagation effects \cite{Harde1991, Harde1991(2)}.
%Determining the temporal localisation of the FID echoes can also be directly used to detect gases without transitioning to the frequency domain \cite{Mittleman1998, Sitnikov2019}.
\begin{figure}[htbp!]
    \centering
    \includegraphics[width=0.7\linewidth]{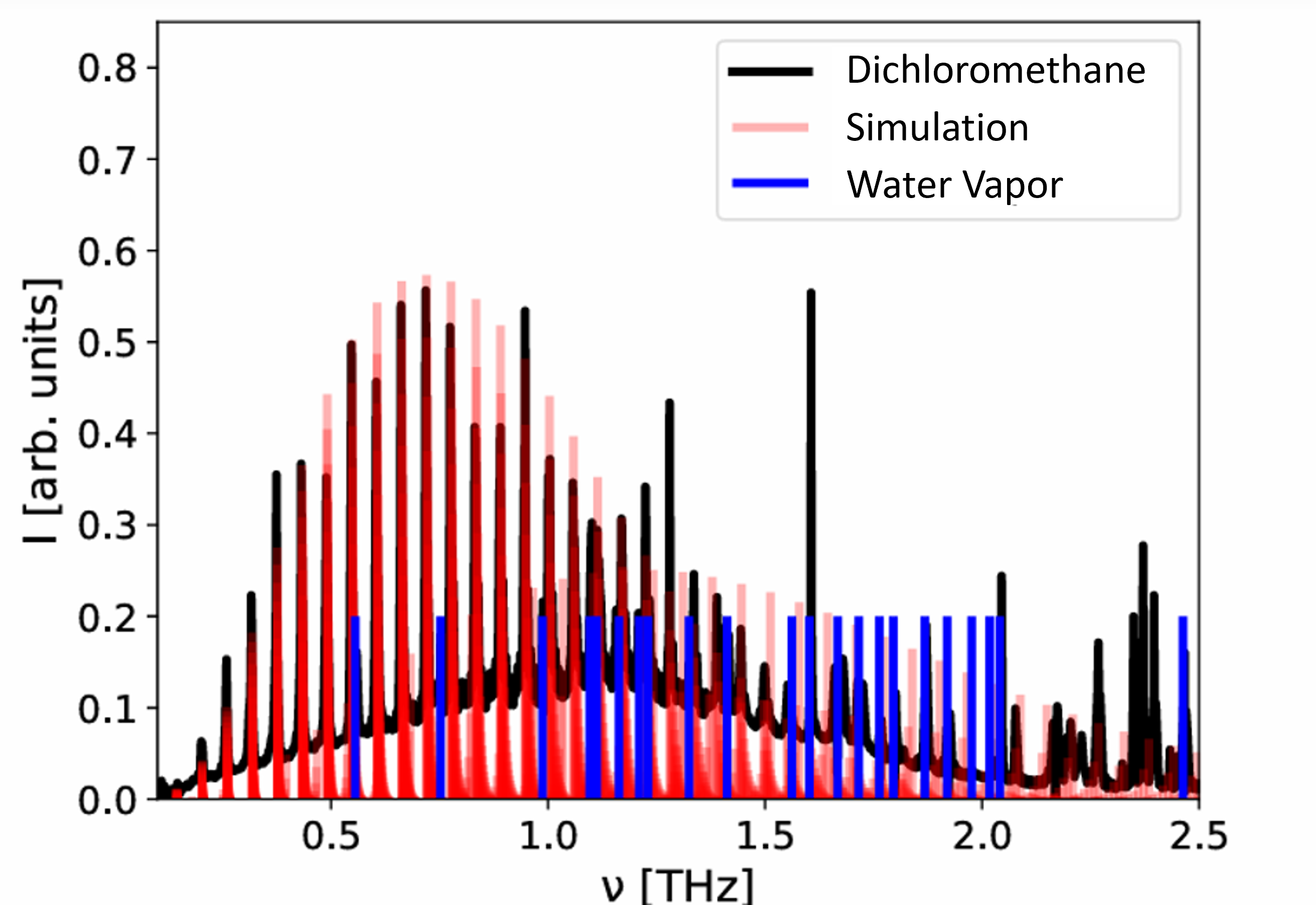}
    \caption{Comparison between DCM’s experimental spectrum (black) and its theoretical calculated rotational transition obtained from a model based on the rigid rotator with only centrifugal corrections (red). Water vapour absorption frequencies are reported for reference (blue). The frequency values have been extracted from the experimental reference spectrum and their amplitude is arbitrarily chosen for graphical representation in order to not cover the sample signal.}
    \label{DCM_sim}
\end{figure}
Moreover, DCM's spectral features are investigated from a theoretical perspective and the results obtained for the experimental data showed a good agreement with the calculated rotational transitions (Figure \ref{DCM_sim}) \cite{Burkhard1959}. The few spectral lines observed experimentally, but not predicted by the simulation are likely to be attributed to roto-vibrational modes not considered in the model based on the rigid rotator with only centrifugal corrections. %A more in-depth study is beyond the scope of this work. 

%\subsection*{Chloroform}

\textbf{Chloroform} has four primary isotopomers; the most abundant one (natural abundance 42.9\%) is an oblate symmetric top molecule with rotational constants A=B=3.302 GHz, C=1.778 GHz \cite{Carpenter1995, Rice2021}. 
\begin{figure} [htbp!]
 \centering
 \includegraphics[width=\linewidth]{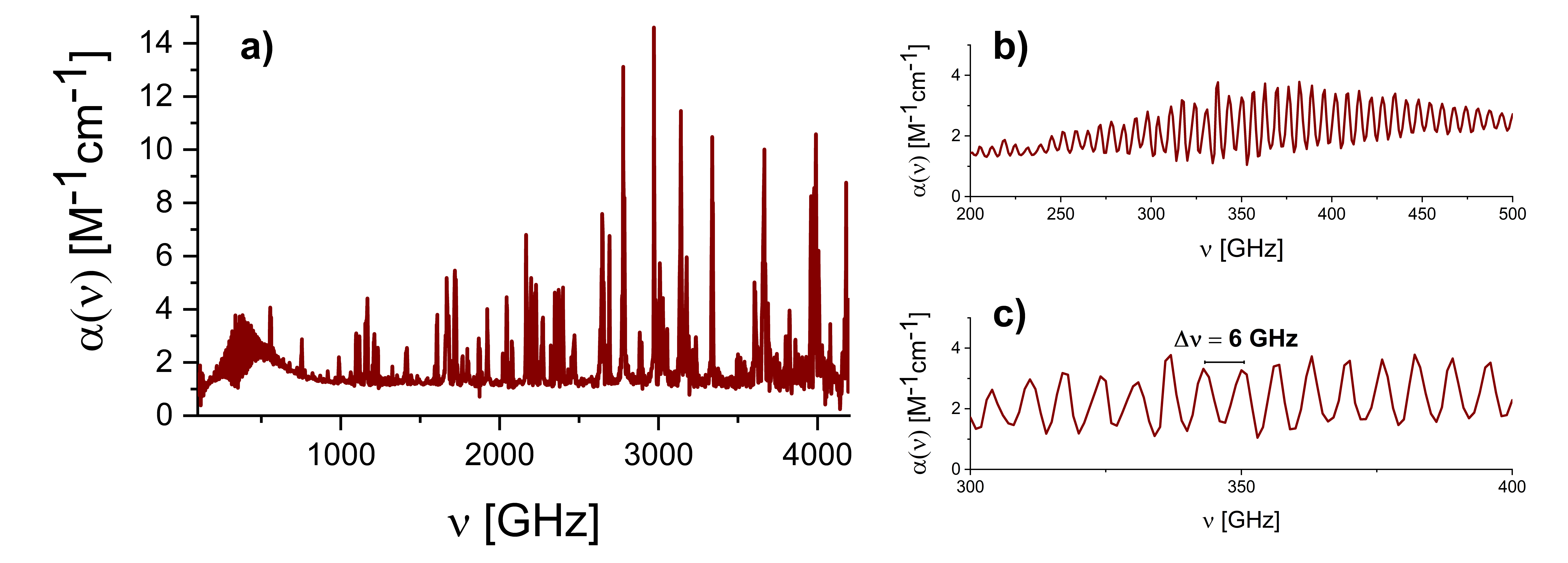}
    \caption{a) THz-TDS spectrum of chloroform (oblate symmetric top molecule with Ray’s parameter $\kappa$=+1) in the spectral range 0.1-4.2 THz; b) Detail of THz-TDS spectrum of chloroform in which the repeating R-branch transitions spaced every $\Delta \nu$=6 GHz in the range 0.2-0.5 THz are evident. The main peaks present above 1.0 THz are associated with water vapour absorptions; c) Detail in the range 0.3-0.4 THz where rotational transitions frequencies are reported.}
    \label{cloroformio_TDS}
\end{figure}
Chloroform investigated with THz-TDS shows strong repeating transitions that are spaced in frequency approximately every 6 GHz, with a FWHM of approximately 3 GHz (Figure \ref{cloroformio_TDS}).
Therefore, the spacing of the peaks associated with R-branch transitions ($\Delta J=+1$) occurs approximately every 2B ($\sim 6.604$ GHz).
%\begin{figure}
   % \includegraphics[width=\linewidth]{Chloroform_alpha_TDS.png}
    %\caption{a) THz-TDS spectrum of chloroform in the spectral range 0.1-4.2 THz; b) Detail of THz-TDS spectrum of chloroform in which are evident the repeating transitions spaced every 6 GHz in the range 0.2-0.5 THz.}
    %\label{cloroformio_TDS}
%\end{figure}
\begin{figure}[htbp!]
    \centering
    \includegraphics[width=0.6\linewidth]{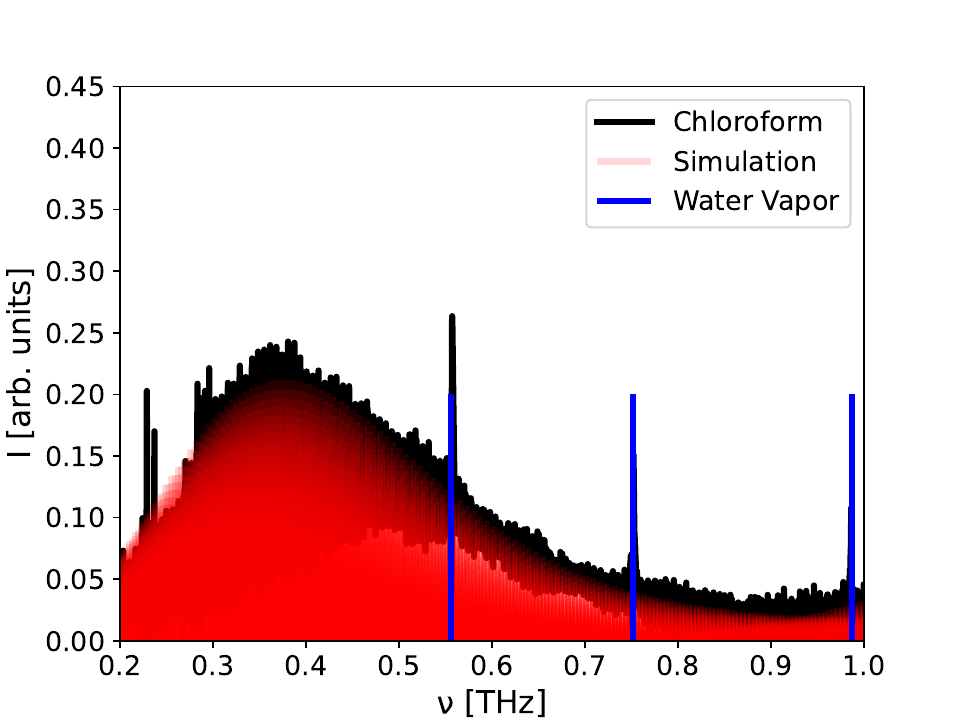}
    \caption{Comparison between chloroform’s experimental spectrum (black) and its theoretical calculated rotational transition obtained from a model based on the rigid rotator with only centrifugal corrections (red). The spacing of the absorption peaks of chloroform (every 6 GHz) is not clearly visible from the graphics but it is confirmed from the theoretical calculations. Water vapour absorption frequencies are reported for reference (blue). The frequency values have been extracted from the experimental reference spectrum and their amplitude is arbitrarily chosen for graphical representation in order to not cover the sample signal.}
    \label{Cl_sim}
\end{figure}
%The temporal profile analysis shows the presence of the weak FID signal separated from the primary THz pulse of around 153 ps.  
%presence of a second pulse separated from the primary THz pulse of around 153 ps.  
%This behaviour is connected to the fact that the FID signal is typically weaker over a long temporal window since it decays exponentially over time \cite{Harde1991, Harde1991(2)}. 
We also compared the experimental spectrum with the theoretical perspective in Figure \ref{Cl_sim} showing a good agreement. Also in this case, the model used has been the rigid rotator with centrifugal corrections.

\subsection{VSLS Mixtures characterization}

\begin{figure}[htbp!]
    \centering
    \includegraphics[width=\linewidth]{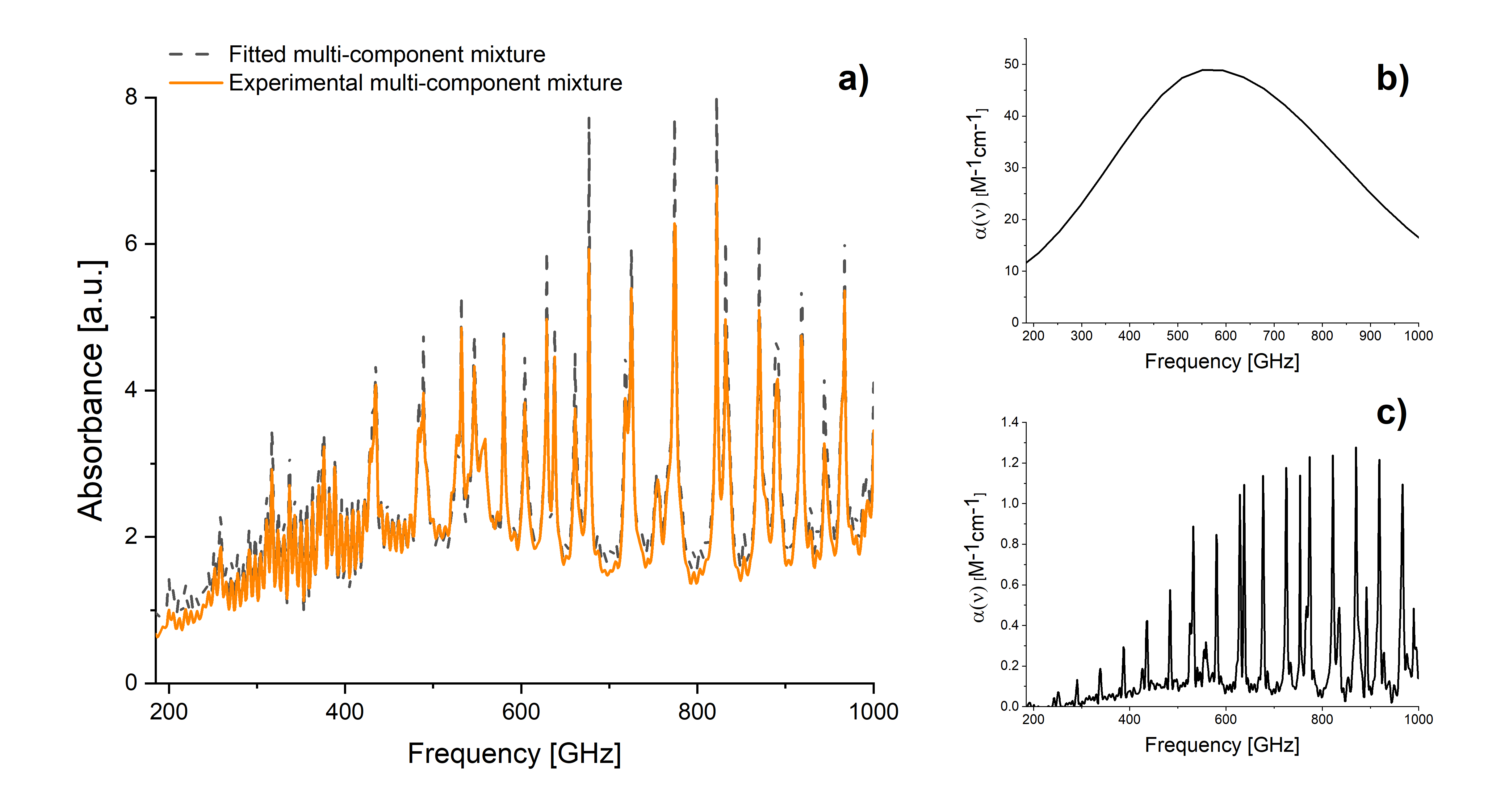}
    \caption{a) Comparison between the experimental absorbance of the multi-component mixture (orange) and the fit obtained with the multiple absorbers approach (black) that allows to retrieve the partial pressure of the single component, as reported in Table 1, in the range 0.2-1 THz where the specific fingerprints of the chemicals investigated are present; b) THz-TDS absorbance spectrum of pure acetone in the range of interest; c) THz-TDS absorbance spectrum of pure methanol in the range of interest}
    \label{fitmix}
\end{figure}

We analyze the THz spectra of the multi-component mixture following the multiple absorbers approach where the interaction between the compounds can be considered negligible and the total spectrum is given by the weighted linear combination of the pure compound's spectra, according to Equation \ref{mix} of the section "Measurement protocol and data analysis". 
For the mixture measurements, we first developed the set-up for \textit{laboratory conditions} when we injected the substances directly into the gas-cell circuit while measuring the partial pressure of the gas with a dedicated pressure gauge and at the same time probing the gas with the THz beam. 
The experimental absorbance fitted with the weighted linear combination of the pure compounds shows an outstanding agreement as shown in Figures \ref{fitmix} and the parameters values of the partial pressures retrieved from the fitting procedure are in optimal agreement with the measurement data as reported in Table \ref{tab1} denoting the capability of the system to give qualitative and also quantitative information.

\begin{figure}[htbp!]
    \centering
    \includegraphics[width=\linewidth]{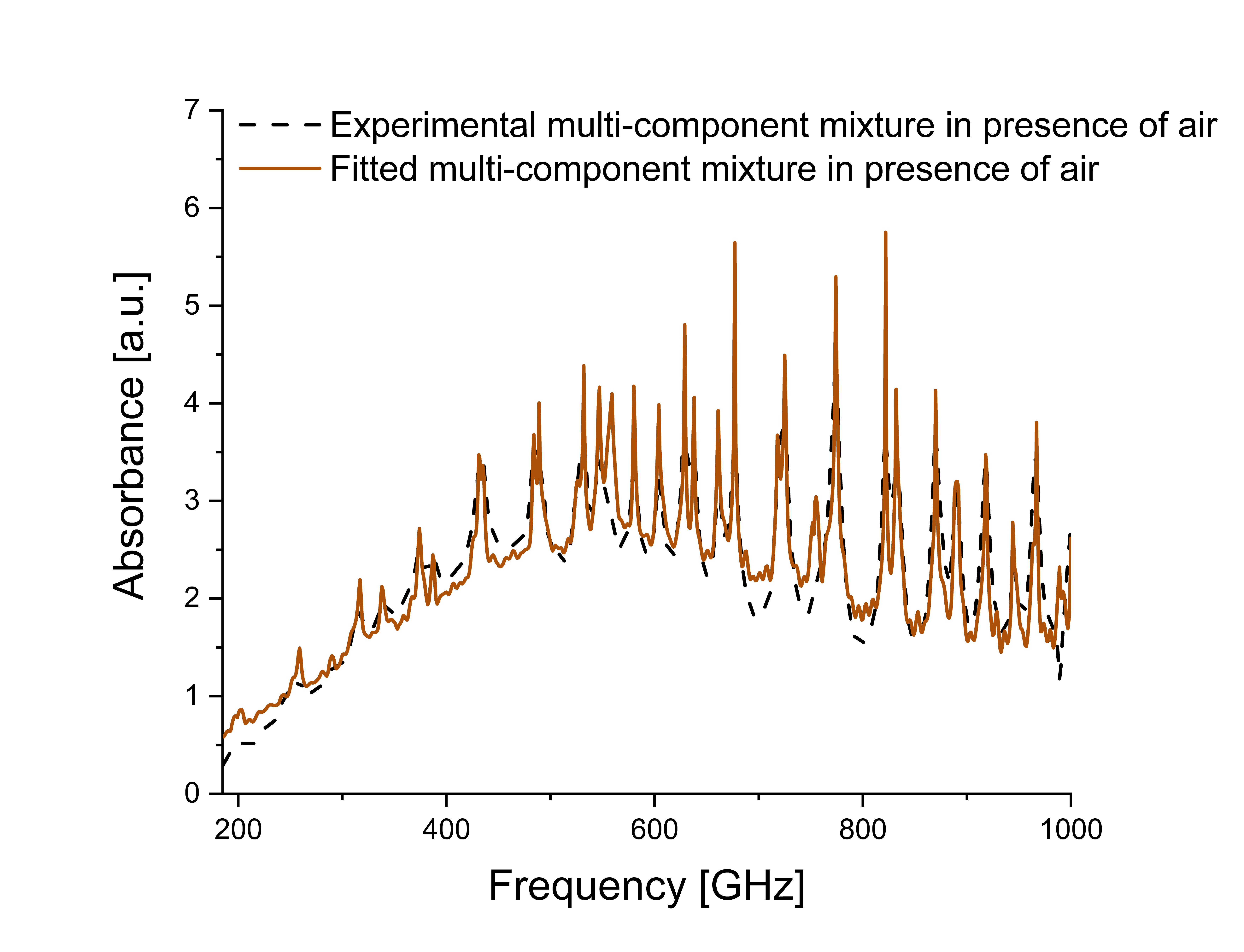}
    \caption{Comparison between the result of the experimental measurement of the multi-component mixture constituted by acetone, methanol, chloroform and dichloromethane (black) dispersed in the atmosphere in a controlled room laboratory condition and then aspired in the main cell through a 5-meter-long pipe (in-field configuration) vs the fit (brown) obtained through the multiple absorbers approach that allows retrieving the partial pressures of the single components ($p_{Chloroform}$=20.7 mbar,  $p_{Dichloromethane}$=23.7 mbar, $p_{Methanol}$=34.4 mbar, and $p_{Acetone}$=37.2 mbar). The range of interest is between 0.2 and 1 THz, where all the specific fingerprints are present.}
    \label{Mix4Air}
\end{figure}

Finally, we modified the prototype for \textit{in-field} measurements ensuring that the ambient atmosphere is drawn directly into the gas cell via a long pipe and a suction system opportunely developed. In this configuration, we diluted the different substances in a different laboratory room atmosphere at a five-meter distance from the set-up.
The experimental absorbance fitted by the weighted linear combination of the pure compounds also in this case shows an outstanding agreement as shown in Figure \ref{Mix4Air} allowing to estimate the following pressures for each compound present in room atmosphere:  $p_{Chloroform}$=20.7 mbar,  $p_{Dichloromethane}$=23.7 mbar, $p_{Methanol}$=34.4 mbar, and $p_{Acetone}$=37.2 mbar.
%in  for each prepared mixture both in laboratory conditions where the samples were injected and with the aspiration system at a distance (Figures \ref{fitmix} and \ref{Mix4Air}). In fact, the resulting spectrum 

For completeness, we also report the measured spectrum of the other gaseous substances employed in the multi-component mixture in Figure \ref{fitmix}b and Figure \ref{fitmix}c. 
As expected, the absorbance features of the two VOCs added to the chlorine-based VSLS agree with previous research works and data \cite{DArco2022, Galstyan2021, Rice2020, Smith2015}.
In fact, acetone shows a single broad absorption centered at 0.56 THz \cite{Galstyan2021} while methanol is characterized by periodic absorption features spaced approximately every 2B=50 GHz \cite{DArco2022}.

%The parameters retrieved from the fitting procedure proved an optimal agreement between experimental and theoretical pressure data which are reported in Table \ref{tab1}.

It is important to remark that the experimental absorbance and the retrieved spectrum obtained as a weighted linear combination of the pure compounds show an outstanding agreement in both \textit{laboratory} and \textit{in-field configuration} (see: Figures \ref{fitmix} and \ref{Mix4Air}); the fit shows the specific absorption features connected to the spectra of the corresponding single compounds and supplies quantitative information on the partial pressure present in the mixture thus proving that the methodology and the prototype for \textit{in-field measurement} that we have rigorously engineered represent a groundbreaking advancement in the gas sensing field at THz frequencies.
The observed data validate the consistency of the presented approach paving the way to an optimized set-up for in-situ investigations.

\begin{table}[htbp!]
\centering
\caption{Measured pressures by gauge and retrieved partial pressures obtained from the fit performed as a weighted linear combination of the pure compounds for the multi-component mixture under the \textit{laboratory configuration}.}
\label{tab1}
\vspace{10pt}
\resizebox{0.7\textwidth}{!}{%
\begin{tabular}{c|c|c|}
\cline{2-3}
                                               & \textbf{Measured  Pressure [mbar]} & \textbf{Retrieved  Pressure [mbar]} \\ \hline
\multicolumn{1}{|c|}{\textbf{Chloroform}}      &     120.5                         &       115.4                     \\ \hline
\multicolumn{1}{|c|}{\textbf{Dichloromethane}} & 47.5                              & 42.8                          \\ \hline
\multicolumn{1}{|c|}{\textbf{Methanol}}        & 67.8                             & 60.4                             \\ \hline
\multicolumn{1}{|c|}{\textbf{Acetone}}         & 36.3                             & 35.8                              \\ \hline
\end{tabular}%
}
\end{table}

\section{Discussion}

In this work, we developed a THz-based prototype set-up which can selectively identify air pollutants like Volatile Organic Compounds (VOCs) and Very short‐lived substances (VSLS). The system is portable and, as shown, it allows to perform measurements at a distance.

We used the prototype to characterise for the first time in literature over a broad THz spectral range (up to 4.5 THz), pure dichloromethane and pure chloroform that are two very short-lived pollutants strongly contributing to ozone depletion thus representing a human and environmental threat. The THz characterisation of these compounds strongly contributes to building a robust spectral database of air contaminants produced by anthropogenic and natural sources or activities. As seen, gaseous fingerprints in THz spectral range are unique for several molecules of interest. The measurements reveal pronounced and distinctive absorption patterns for the halogenated hydrocarbon compounds, providing valuable insights into their spectroscopic behaviour.
For dichloromethane, we evidenced the presence of a sequence of terahertz pulses in the time domain separated approximately every 17 ps, corresponding to the frequency separation between adjacent rotational lines of 57 GHz.\\
The chloroform's spectrum highlighted the presence of specific absorption features approximately every 6 GHz.\\
Then we study the optical response of a multi-component mixture achieved with the two aforementioned chlorine-based compounds mixed with two widely distributed volatile pollutants (acetone and methanol). For the first mixture measurements, we developed the set-up specifically for \textit{laboratory conditions} in which the substances are directly injected into the gas-cell circuit under vacuum conditions. Finally, for the \textit{in-field condition}, we modified the prototype to ensure that the ambient atmosphere in a controlled room, where the gases are dispersed, is drawn directly into the gas cell via a five-meter-long pipe and a suction system opportunely developed. The good matching of the mixture spectra, in both \textit{laboratory} and \textit{in-field conditions}, with the fitting model based on the multiple absorbers approach, demonstrates that the prototype together with the analysis model employed in this work can simultaneously identify and quantify single components in the atmosphere. 
For this proof of concept, in the measurements conditions, the limit of detection with safe margin is about 0.5 microliters of compounds that completely evaporate in the volume of the cell.

The findings presented herein establish a promising foundation for the development and application of an efficient, portable THz-based sensor for the remote detection of multi-component environmental contaminants. The successful adaptation of the prototype for in situ measurements, achieved through the integration of a custom-engineered suction system and a five-meter intake line, enabled direct sampling of ambient air from a controlled environment to the system's control unit. The relatively lightweight and compact design of the control and measurement unit facilitates its transport and operation in field environments without significant logistical constraints.
Further research will concern the possibility of engineering the proposed prototype to maximize its sensibility to a wide range of gases while keeping the compactness and portability of the system. 
Moreover, a dedicated iterative algorithm encompassing a machine-learning approach can be implemented to further improve the goodness of the fitting of the experimental data thus obtaining higher precision for the retrieved partial pressure values.
In this work, we selected THz-TDS specifically to develop a prototype for broadband gas detection to observe more chemicals with short acquisition time and avoid peak overlapping in complex mixtures \cite{Tyree2022}. 
Nevertheless, the presented prototype can be implemented also with THz-CW or FTIR-THz technologies.
Further studies will be conducted to develop complementary in-field prototypes based on high-frequency resolution terahertz continuous wave spectroscopy (THz-CW) where the frequency resolution (in the order of MHz) can be exploited for gas sensing to evidence small spectral shifts, peak broadening or detect chemicals requiring a frequency resolution below 500 MHz, which is the highest resolution of the THz-TDS system described \cite{Cheville1995, Cheville1999, Matton2006, Bigourd2007, Cazzoli2008, Cazzoli2009, Svelto2010, Ren2010}.
It is worth mentioning for variations of the real ambient conditions expected in a real case scenario as proposed in this work, peak shifts and line broadening are negligible compared to the system resolution as discussed in previous works \cite{Cheville1995, Cheville1999, Matton2006, Bigourd2007, Cazzoli2008, Cazzoli2009, Svelto2010, Ren2010}.
Such developments are currently being carried out to build an integrated unmanned aircraft system-terahertz spectrometer (UAS-THz) representing a highly valuable remote sensor for atmospheric pollutants especially in situations where distance, obstacles, or atmospheric conditions prevent a direct observation.

\section{Methods}

\subsection{Layout of the prototype}
\label{gascell}

\begin{figure}[htbp!]
    \centering
    \includegraphics[width=\linewidth]{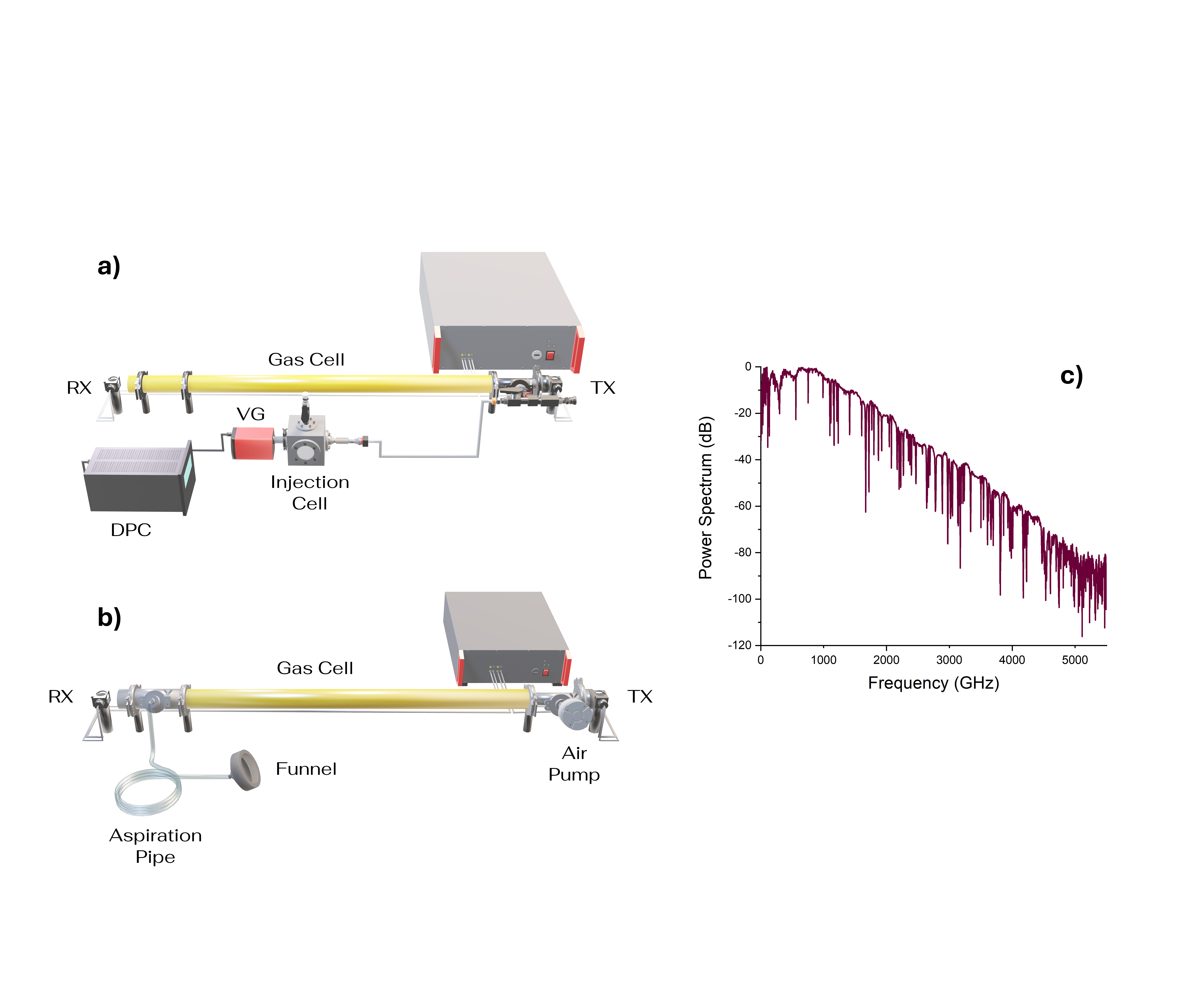}
    \caption{a) Schematic representation of the experimental set-up based on THz-TDS for gas analysis for the \textit{laboratory configuration}. TX and RX are the two photoconductive antennas that act as transmitter and receiver of THz radiation, respectively. VG is the vacuum gauge associated with the precision digital pressure controller (DPC); b) Schematic representation of the experimental set-up based on THz-TDS for gas analysis for the \textit{in-field configuration} with the prototype assembled with the air pump and the 5 m long aspiration pipe that ends with the funnel for the aspiration of the target inside the cell; c) Power spectrum (dB) of the gas cell in presence of ambient air closed by two Teflon windows where the bandwidth reaches 5.5 THz, the peaks present are related to water vapour absorption.}
    \label{setup}
\end{figure}

This work employed a custom-designed gas cell to investigate the interactions between THz radiation and the selected gases under controlled conditions (schematically represented in Figure \ref{setup}a). 

The main gas cell is 1.10 meters long and comprises a cylindrical chamber (with a diameter of 21.5 mm) constructed from brass. 
The cylindrical gas cell is equipped with two Teflon windows (with a thickness of 4.5 mm each) that are highly transparent in the THz range up to approximately 5.5 THz \cite{Naftaly2015}. The windows are placed perpendicular to the THz radiation direction. 

The prototype presented is combined with a terahertz time-domain spectrometer (TeraFlash pro system, TOPTICA Photonics AG, Germany). The system consists of an all-fiber-based femtosecond laser system (FemtoFErb 1560, Toptica Photonics). %The laser pulses are emitted at a wavelength of 1560 nm and the repetition rate is 80 MHz. The pulses have a halfwidth of approximately 80 fs.\\
%The output beam is divided by a 50:50 fiber splitter in the emitter and detector branches. 
%The optical pulses in both arms are guided to fiber-coupled, InGaAs-based 
The optical pulses are converted into THz pulses through a photoconductive antenna, which serves as the emitter (indicated as TX in Figure \ref{setup}a and Figure \ref{setup}b); another photoconductive antenna acts as detector (RX).
The THz beam is collimated by a gold-coated off-axis parabolic mirror (OAP, RFL=25.4 mm) placed inside a 30 mm mirror mount cage system; the transmitted beam is then focused to the RX by an OAP (RFL=25.4 mm) placed inside the mount cage system
%and detector. 
%The emitter converts the laser light into a terahertz pulse. 
A fast and highly precise delay provides the time variation between the emitter and receiver branches required for sampling the terahertz pulse. A peak dynamic range of approximately 100 dB is reachable with 1000 averages and the corresponding usable bandwidth reaches 6 THz when the Teflon windows are not present, otherwise, the bandwidth reaches 5.5 THz (Figure \ref{setup}c).

In what is addressed through the paper as \textit{laboratory configuration}, the long gas cell is connected to a secondary injection cell (Figure \ref{setup}a) for the introduction of the compounds in their liquid phase, ensuring a closed and controlled experimental environment. The injection cell with a length of about 10 cm is much smaller than the main one.  The liquid samples are injected into the injection cell through calibrated gas syringes. We monitor and control the gas pressure, by a vacuum gauge (VG) associated with a precision digital pressure controller (DPC, CCT 361 PT R50 130 and Omnicontrol 200, Pfeiffer Vacuum GmbH, Germany).

%The measurements are conducted at room temperature. At first, we measure the spectra of the empty gas cell (under vacuum at P $\approx$ 90 hPa) and we use it as a reference. This also allows the minimization of the water vapour contribution to the resulting spectra. Then we inject the gas and we probe it with the THz beam. The reference spectra are collected each time before the analysis of the samples of interest, which is performed once a steady pressure is reached.
%equilibrium (liquid-vapor) is reached ( this condition is monitored through the pressure sensor).\\

In what is addressed through the paper as \textit{in-field configuration}, we connected the main cell directly to an aspiration system consisting of an air pump and a 5 m long aspiration pipe (with an average inner diameter of 6.5 mm) which ends with funnel with a wide opening through which the air is drawn inside the main cell (as schematically reported in Figure \ref{setup}b); the injection cell and the vacuum gauge system (Injection Cell, VG and DPC, as indicated in Figure \ref{setup}a) are now uncoupled from the system making it more compact and lightweight. Then we proceeded to perform measurements of the mixture dispersed in a second laboratory room five meters apart. 
%atmosfpere 5selected compounds with air (RH=30\%). Thus, the components of the multi-component mixture were drawn into the gas cell using the pumping system connected to a 5-meter-long tube. This setup facilitated the efficient extraction of the compounds from a considerable distance.
All experiments are conducted in multiple replicas to ensure repeatability. 

\subsection{Samples}
Dichloromethane ($CH_2Cl_2$) with a purity $\geq$ 99.8\% (CAS No.: 75-09-2) and chloroform ($CHCl_3$) with a purity $\geq$ 99\% (CAS No.: 67-66-3) are purchased from Sigma-Aldrich. 
The liquid compound is injected into the secondary inlet cell connected to the main measurement gas cell (1.10 m).
For the multi-component mixture, gaseous substances belonging both to VSLS and VOCs are selected.
Specifically, we investigated a gas mixture prepared via injecting: dichloromethane, chloroform, acetone, and methanol.
Acetone ($CH_3COCH_3$) with a purity of $\geq$ 99.5\% (CAS No.: 67-64-1) and methanol ($CH_3OH$) with a purity of $\geq$ 99.8\% (CAS No.: 67-56-1) are also purchased from Sigma-Aldrich.
For completeness, experimental spectra of acetone and methanol are reported in Figure \ref{fitmix}b and Figure \ref{fitmix}c.
The multi-component mixture is prepared by injecting each single compound at a time and measuring their relative pressure. The injection sequence is chloroform, methanol, dichloromethane, and acetone.
Acetone and methanol have already been characterized in the THz spectral region and present specific fingerprints at these frequencies \cite{DArco2022, Galstyan2021, Rice2020, Smith2015}.
In this context, we also characterized with our prototype the known compounds and we built a spectral library to use for the data analysis. 

\subsection{Measurement protocol and data analysis}

For the gaseous compounds characterization (single pure substances or their multi-component mixture), performed in the \textit{laboratory configuration}, the cell circuit is pumped under vacuum (P $\approx$ 90 hPa) and the reference spectrum is acquired; then, the samples of interest are injected into the cell in their liquid form, and the gaseous phase is probed with THz radiation propagated in the main long cell.  The reference spectra are collected each time before the analysis of the samples of interest, which is performed once a steady pressure is reached. \\
In order to simulate practical measurements for \textit{in-field conditions}, the set-up configuration is modified as described in the section "Layout of the prototype"; the polluted ambient air at the location of interest, where we dispersed the substances in a controlled atmosphere of a laboratory room, is aspired through the 5-meter tube into the main gas cell where it is probed with the THz beam. %y exploiting the same configuration, the samples of interest are then collected into the main cell and investigated with THz radiation. 
In this condition, we used as a reference the spectrum of non-polluted ambient air (P $\approx$ 1013 hPa) aspired with the 5-meter tube from another location.

All the measurements are conducted at room temperature and recorded in the time domain in a time window of 200 ps which corresponds to a resolution of 5 GHz. The corresponding frequency spectra are obtained from the temporal waveform by numerical Fourier transform \cite{Naftaly2015}. 
We applied zero-padding to the time domain data to obtain a 1 GHz resolution \cite{Walker2012, Withayachumnankul2014}. We analyzed the raw absorbance spectra in MATLAB (ver. 2019a, MathWorks Inc., USA) by applying a Savitzky-Golay (S-G) filter with a quadratic polynomial order of 55 points  \cite{Liu2018, Qiu2024}.

%The spectra of the empty gas cell (under vacuum at P $\approx$ 90 hPa) and in the presence of ambient air (P $\approx$ 1013 hPa) are acquired and used as a reference for the laboratory and in-field configuration, respectively. This also allows the minimization of the water vapour contribution to the resulting spectra. Then the gaseous compounds are injected or aspirated into the gas cell and probed with the THz beam. The reference spectra are collected each time before the analysis of the samples of interest, which is performed once a steady pressure is reached for the laboratory configuration. The measurements are conducted at room temperature.
%Before Fourier transforming, time-domain data were zero-padded  thus obtaining aw 1 GHz resolution.
%thus, they were extended by adding a string of zero values at the end of the waveforms \cite{Walker2012, Withayachumnankul2014}.
%THz denoising procedures to remove unwanted oscillations without compromising the goodness of the data

The sample's transmittance is measured as follows:
\begin{equation}
\centering
T\left(\nu\right)=\frac{\left(I_{sample}\left(\nu\right)\right)}{\left(I_{reference}\left(\nu\right)\right)}    
\end{equation}

Where $I_{sample}$ and $I_{reference}$ are the energy spectral densities of the sample and the reference, respectively.
Subsequently, according to the Beer–Lambert Law, the experimental absorbance, $Absorbance(\nu)$, is determined as follows:

\begin{equation}
\centering
Absorbance(\nu)=-log\frac{I_{sample}}{I_{reference}}=-log(T(\nu))   
\end{equation}

The molecular absorption coefficient $\alpha_M(\nu)$ is obtained by combining the Lamber-Beer law with the perfect gas law as:
\begin{equation}
\centering
\alpha_M (\nu)\ =\frac{Absorbance(\nu)}{C\cdot d}=\frac{Absorbance(\nu)\cdot R\cdot T}{P \cdot d}\     
\label{alpha}
\end{equation}

where $C$ is the molar concentration of the gas in the cell [M], $d$ [cm] is the optical path of the
gas cell, $R$ is the ideal gas constant [$L\cdot Pa \cdot K^{-1}\cdot mol^{-1}$], $T$ is the temperature expressed in Kelvin [K] and $P$ [Pa] is the measured pressure. 

The simulations have been performed following three main references. The centrifugally distorted rotational levels have been calculated, up to sextic terms in the rotational quantum numbers \cite{Martin1991}. The higher-order centrifugal coefficients have been found through a fit on the experimental data, and their order of magnitude is consistent with the scaling reported in previous research works \cite{Martin1991, Watson1968}. For the calculation of the envelopes of the emission bands at finite temperatures, we have referred to the work of Gerhard and Dennison \cite{Gerhard1933}, where the degeneracy and occupancy levels have been calculated for each of the molecular states involved in the radiation emission. 

To analyze the multi-component mixture retrieving the partial pressure of the gasses, assuming that there is no inter-gas interaction between the components, we exploit a model based on the multiple absorbers approach to fit the experimental data \cite{DArco2022, Moffa2024, Moffa2024(2)} (we employed the algorithm "lsqcurvefit", MATLAB ver. 2019a, MathWorks Inc., USA) , where the concentration of each compound is retrieved from a weighted linear combination: 

%\begin{equation}
 %   Absorbance_{Retrieved}(\nu)=\alpha_{DCM}(\nu)\cdot a+\alpha_{Cloroform}(\nu)\cdot b%+\alpha_{Acetone}(\nu)\cdot c+\alpha_{Methanol}(\nu)\cdot d
   % \label{mix}
%\end{equation}

\begin{equation}
     Absorbance_{Retrieved}(\nu)=\sum_{i=1}^{n} \alpha_{i}(\nu)\cdot x_i
     \label{mix}
\end{equation}

where $\alpha_{i}(\nu)$ are the molecular absorption coefficients as a function of frequency (expressed in $M^{-1}cm^{-1}$) of the pure compounds $\alpha_{DCM}(\nu)$, $\alpha_{Cloroform}(\nu)$, $\alpha_{Acetone}(\nu)$, and $\alpha_{Methanol}(\nu)$ and $x_i$ represent the fitting parameter corresponding to the retrieved partial pressure of each compound (expressed in mbar).
The procedure allows the exclusion of the compounds that are not present in the mixture (specifically, the fitting procedure will give a coefficient connected to the partial pressure equal to 0).

\bibliography{biblio}

\section*{Acknowledgements}

\noindent This work was carried out thanks to "PRIN 2022: TREX a prototype of a portable and remotely controlled platform based on THz technology to measure the one health vision: environment, food, plant health, security, human and animal health" funded by the European Union - Next Generation EU (CUP B53D23013610006 - Project Code 2022B3MLXB PNRR M4.C2.1.1). This work has been supported by PNRR MUR project
CN00000022 “Agritech” spoke 9. This research was conducted in the framework of "STORM - Sensori su sistemi mobili e remoti al Terahertz PNRM a2017.153 STORM" funded by Ministero della Difesa and "R-SET: Remote sensing for the environment by THz radiation" Large Research Projects of Sapienza, University of Rome. This work was also supported by Sapienza competitive grants: Grandi Attrezzature Scientifiche (2018) titled "SapienzaTerahertz: THz spectroscopic image system for basic and applied sciences".  This work was supported by Sapienza Large Projects Research Call 2023 titled "TforCH: R\&D on the potentiality of THz radiation for Cultural Heritage". The authors would like to thank "CSN5 Grants INFN Roma 1" for their contribution to this work.

\section*{Author contributions statement}

\noindent Conceptualization, C.M. and M.P.; methodology, C.M. and M.P.; software, C.M., A.C., M.C. and M.P.; validation, C.M., A.C., C.M.,  V.M.O., D.F., F.Jr.P.M., L.G., M.M., G.Z., M.R., L.M., and M.P.; formal analysis, C.M., A.C. and M.P.; investigation, C.M., and M.P.; resources, M.M., M.C., L.G., M.M., G.Z., M.R., L.M., and M.P.; data curation, C.M., A.C., and M.P.; writing---original draft preparation, C.M. and M.P.; writing---review and editing, all the authors; visualization, C.M. and A.C.; supervision, A.C., L.G., M.M., G.Z., M.R., L.M., and M.P.; project administration, C.M., and M.P.; funding acquisition, L.G., M.R., L.M., and M.P. \\
All authors have read and agreed to the published version of the manuscript.

\section*{Additional information}

\noindent \textbf{Competing interests}: The authors declare no competing interests. 
\\
\textbf{Data availability}: The datasets used and/or analysed during the current study are available from the corresponding author on reasonable request.

\end{document}